\documentstyle[12pt,psfig]{article} 
\let\section=\subsection     \let\subsection=\subsubsection                
\newcommand{\puncspace}{\ifmmode\,\else{\ifcat.\C{\if.\C\else%
\if,\C\else\if?\C\else\if:\C\else\if;\C\else\if-\C\else%
\if)\C\else\if/\C\else\if]\C\else\if'\C%
\else\space\fi\fi\fi\fi\fi\fi\fi\fi\fi\fi}%
\else\if\empty\C\else\if\space\C\else\space\fi\fi\fi}\fi}%
\newcommand{\SP}{\let\\=\empty\futurelet\C\puncspace}
\newcommand{\nonrot}{{\rm 0}}
\newcommand{\dash}{\hbox{--}}
\newcommand{\Hz}{{\hbox{Hz}}\SP}
\def\fu#1{\leavevmode\hbox{4U~#1}\SP}
\def\rxte{\leavevmode{\it RXTE}\SP}
\newcommand{\gta}{\ifmmode {\mathbin{\lower 3pt\hbox   
    {$\,\rlap{\raise 5pt\hbox{$\char'076$}}\mathchar"7218\,$}}}
    \else {${\mathbin{\lower 3pt\hbox
    {$\rlap{\raise 5pt\hbox{$\char'076$}}\mathchar"7218\,$}}}
    $}\fi}

\begin{document}
\begin{center}
   {\large \bf CONSTRAINTS ON NEUTRON STAR MATTER} \\[2mm]
   {\large \bf FROM KILOHERTZ QPOs}\\[5mm]
   F.~K. LAMB$^1$, M.~C. MILLER$^2$, and D. PSALTIS$^3$ \\[5mm]
   {\small \it  $^1$University of Illinois at Urbana-Champaign \\     
     Department of Physics and Department of Astronomy \\
     1110 W. Green St., Urbana, IL  61801, USA \\ [1mm]
   $^2$University of Chicago, Department of Astronomy and Astrophysics \\
     5640 S. Ellis Avenue, Chicago, IL  60637, USA \\ [1mm]
   $^3$Harvard-Smithsonian Center for Astrophysics \\
     60 Garden St., Cambridge, MA 02138, USA \\[8mm] }
\end{center}

 \begin{abstract}\noindent
 One of the most dramatic discoveries
made so far with the {\em Rossi X-Ray
Timing Explorer\/} is that many accreting
neutron stars with weak magnetic fields
generate strong, remarkably coherent,
high-frequency X-ray brightness
oscillations.
 The \hbox{$\sim$325--1200~Hz}
quasi-periodic oscillations (QPOs)
observed in the accretion-powered
emission are almost certainly produced
by gas orbiting very close to the
stellar surface and have frequencies
related to the orbital frequencies of
the gas. 
 The \hbox{$\sim$360--600~Hz} brightness
oscillations seen during thermonuclear
X-ray bursts are produced by one or two
hotter regions on the stellar surface and
have frequencies equal to the stellar
spin frequency or its first overtone.
 Measurements of these oscillations are
providing tight upper bounds on the
masses and radii of neutron stars, and
important new constraints on the
equation of state of neutron star matter.
 \end{abstract}

\section{Introduction}

Since the birth of X-ray astronomy 35
years ago, scientists have sought to use
the X-radiation that comes from near the
event horizons of black holes and the
surfaces of neutron stars to probe
quantitatively the strong gravitational
fields near these objects and to
determine the fundamental properties of
dense matter (see, e.g.,
\cite{ELP86,LP79}). The {\em Rossi X-Ray
Timing Explorer\/} (\rxte), which was
launched on December 30, 1995, was
specially designed to have the large
area, microsecond time resolution, high
telemetry bandwidth, and pointing
flexibility needed to address these
questions (see \cite{Swank95}). With
\rxte, strong, high-frequency X-ray
brightness oscillations have been
discovered from at least two black holes
(see \cite{McClintock98}) and
sixteen neutron stars (see
\cite{MLP98a,vdK98}). As a result, we
appear to be on the threshold of
achieving this decades-old goal of X-ray
astronomy. Here we focus on the
oscillations discovered in neutron stars
and their implications for the
properties of these stars and for
neutron star matter.

\section{High-Frequency Brightness
Oscillations}

\subsection{Observed Properties}

High-frequency X-ray brightness
oscillations are observed both in the
transient X-ray emission produced during
type~I (thermonuclear) X-ray bursts (see
\cite{SSZ98}) and in the persistent,
accretion-powered X-ray emission (see
\cite{vdK98}). 

The high-frequency oscillations observed
during X-ray bursts have frequencies in
the range 360--600~Hz and rms amplitudes
at least as high as $\sim$35\%
\cite{SZS97}. Only a single oscillation
is observed during a burst, the
oscillation appears to be highly coherent
during the burst decay (see, e.g.,
\cite{SMB97}), and the frequency of the
oscillations produced by a given star is
always the same (measurements of the
burst oscillations in \fu{1728$-$34}
over about a year show that the
timescale of any variation in the
oscillation frequency is $\gta 3000$~yr
\cite{Strohmayer97}). The evidence is
compelling that the oscillations seen
during X-ray bursts are produced by
regions of brighter X-ray emission that
rotate with the star, and that the
frequency of the burst oscillations is
the stellar spin frequency or its first
overtone (see \cite{MLP98a,SZS97}). The
spin frequencies of these neutron stars
appear to be in the range 250--350~Hz
\cite{MLP98a}.

The kilohertz quasi-periodic oscillations
(QPOs) observed in the persistent X-ray
emission have frequencies in the range
325--1200~Hz, rms amplitudes as high as
$\sim$15\%, and quality factors
$\nu/\delta\nu$ as high as $\sim$200
\cite{vdK98}. Two kilohertz QPOs are
commonly observed simultaneously in a
given source (see Fig.~1). Although the
frequencies of the two QPOs vary by
hundreds of Hertz, the frequency
separation $\Delta\nu$ between them
appears to be nearly constant in almost
all cases (see \cite{vdK98}). The
frequency separation of the two
kilohertz QPOs observed in a given star
is closely equal to the spin frequency
of the star inferred from its burst
oscillations (see
\cite{MLP98a,Strohmayer96}).

 \begin{figure}[t!]  
 \begin{center}
 \begin{minipage}{9cm}
 \vglue-0.2truecm
\psfig{file=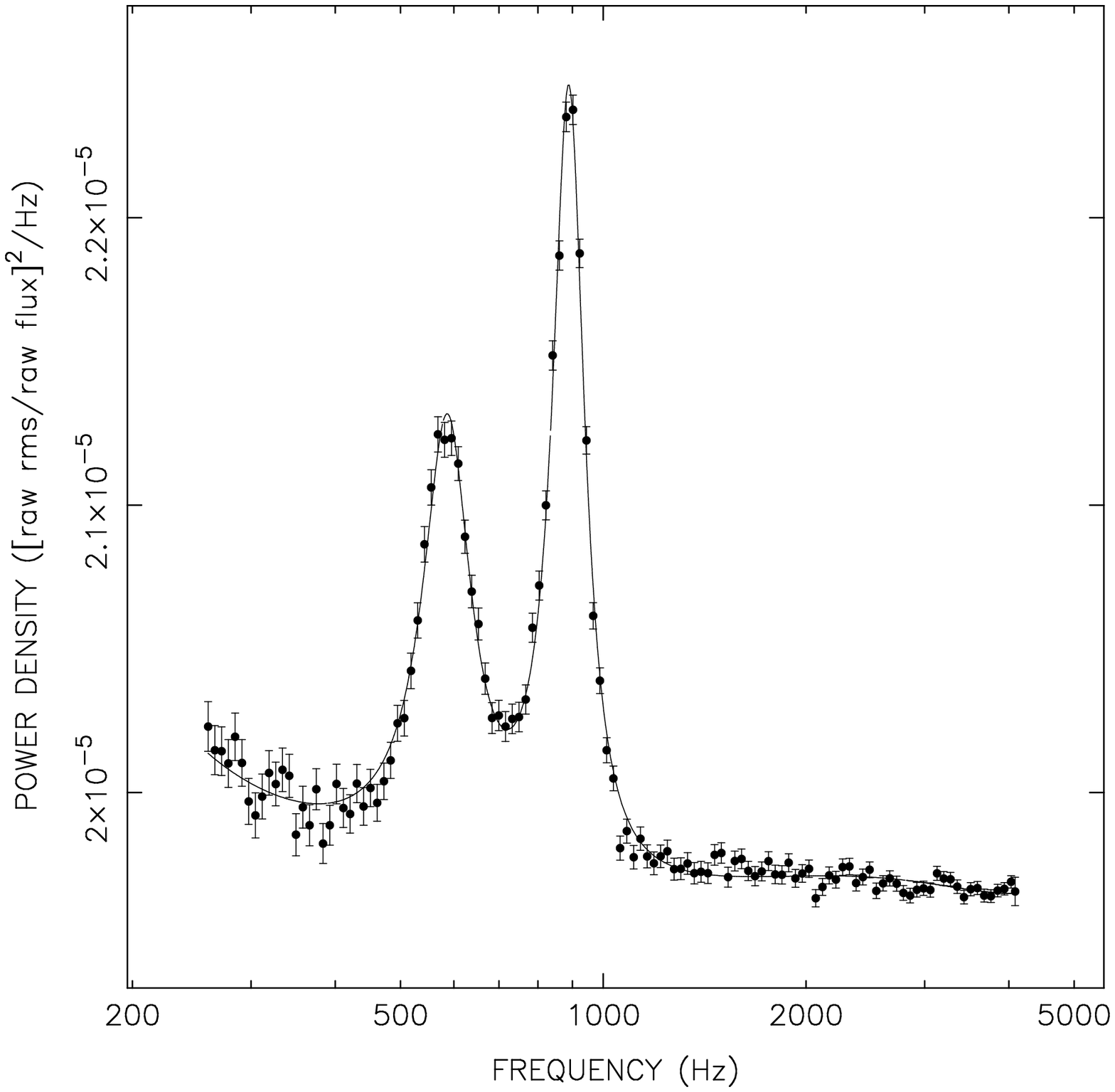,height=7.5truecm}
 \end{minipage}
 \hglue 0.5truecm
 \begin{minipage}{4.5cm}
 \vglue-0.5truecm
 \baselineskip12pt
 {\begin{small}
 Fig.~1. Power spectrum of Sco~X-1 X-ray
brightness variations, showing the two
simultaneous kilohertz QPOs that are
characteristic of the kilohertz QPO
sources.
 The QPOs observed in Sco~X-1 are two of
the weakest kilohertz QPOs detected,
with rms amplitudes $\sim$1\%.
 The continuum power density is
consistent with that expected from
photon counting noise alone.
 From \cite{vdK97}. \newline
 \end{small}}
 \end{minipage}
 \end{center}
 \end{figure}

\subsection{Origin of Kilohertz QPOs}

Although other types of models have been
suggested \cite{Klein96,TM97}, the
fact that the separation $\Delta\nu$
between the frequencies of the two
kilohertz QPOs observed from a given star
is closely equal to the spin frequency of
the star inferred from its burst
oscillations is very strong evidence in
favor of beat-frequency models
\cite{vdK98}. In these models, the
frequency of the higher-frequency QPO is
the Keplerian frequency at a special
radius and the frequency of the
lower-frequency QPO is the difference
between the higher frequency and the spin
frequency $\nu_{\rm spin}$ of the
neutron star. The magnetospheric
beat-frequency model, which was
developed to explain the single,
$\sim$15--60~Hz X-ray brightness
oscillations discovered in the Z sources
a decade ago (see \cite{Lamb91}), has
been discussed \cite{Strohmayer96} as a
possible explanation for the kilohertz
QPOs. However, it is difficult to
explain many basic features of the
kilohertz QPOs using this model,
including why there are {\em two\/}
such QPOs (see \cite{MLP98a}).

The most fully developed and successful
model of the kilohertz QPOs is the
so-called sonic-point beat-frequency
model \cite{MLP98a}. This model is based
on previous work \cite{ML93,ML96} which
showed that the drag force produced by
radiation from a central star can
terminate a Keplerian disk flow near the
star. In the sonic-point model, some
accreting gas spirals inward in nearly
circular orbits until it is close to the
neutron star, where radiation forces or
general relativistic effects cause a
sudden increase in the inward radial
velocity. The radius at which this
occurs is conveniently referred to as
the sonic radius, even though the
transition to supersonic flow is not
directly relevant in this model.

The sharp increase in the inward
velocity is usually caused by the drag
exerted on the orbiting gas by
radiation from the star, but may instead
be caused by general relativistic
corrections to Newtonian gravity, if the
gas in the Keplerian flow reaches
the innermost stable circular orbit
without being significantly affected by
radiation. Gas streams inward from
density fluctuations (clumps) orbiting
near the sonic radius along tightly
spiraling trajectories like that shown in
Figure~2a, generating a more open spiral
density pattern like that shown in
Figure~2b. This pattern rotates around
the star with the Keplerian orbital
frequency at the sonic point, $\nu_{\rm
Ks}$. Collision of the denser gas from
the clumps with the stellar surface
creates beams of brighter X-ray
emission, like that indicated by the
white dashed lines in Figure~2b. These
beams move around the star's equator,
generating a quasi-periodic brightness
oscillation with frequency $\nu_{\rm
Ks}$. Accreting gas is funneled to
certain parts of the stellar surface by
the star's weak magnetic field,
producing weak X-ray beams that rotate
{\em with the star}. These beams are
overtaken by a given orbiting clump once
each beat period, so the inward
mass flux from the clumps, and hence the
accretion luminosity, varies at the
sonic-point beat frequency
\hbox{$\nu_{\rm Ks}-\nu_{\rm spin}$}.

 \begin{figure*}[t!] 
 \label{spirals}
 \begin{center}
 \begin{minipage}{13cm}
 \vglue-2.7truecm
 \centerline{\hskip0.7truecm
\psfig{file=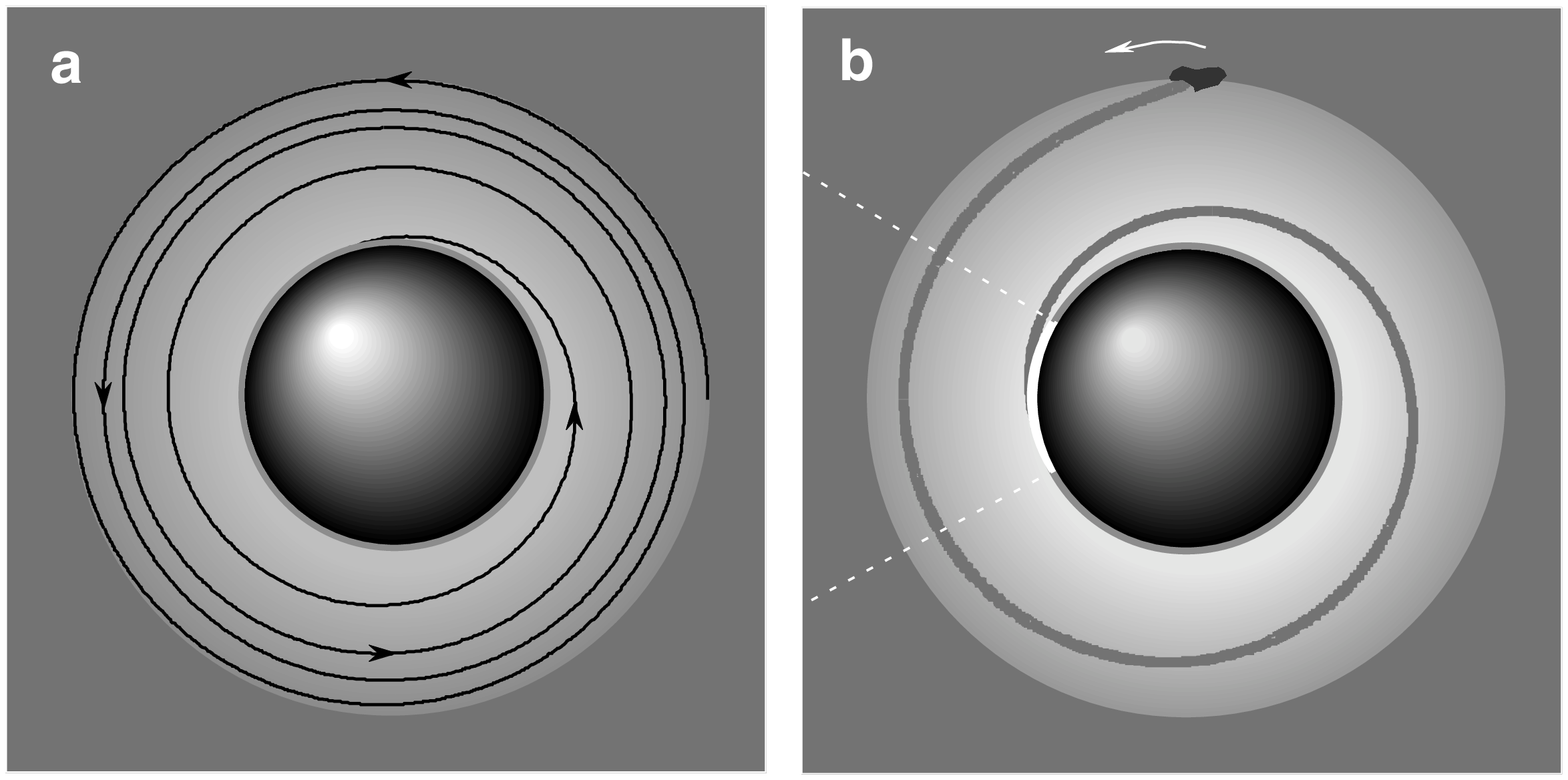,angle=90,height=12.0truecm}}
 \vglue-3.2truecm
 {\begin{small}
 Fig.~2. View of the neutron star and
accretion disk along the rotation axis of
the disk, which is rotating
counterclockwise in this view.
  (a)~Spiral trajectory followed by a
single element of gas as it falls
supersonically from the sonic radius to
the stellar surface.
  (b)~Spiral pattern of higher gas
density formed by gas streaming inward
along spiral trajectories with the shape
shown in (a), from a clump orbiting near
the sonic radius.
 The trajectory and density pattern are
from general relativistic gas dynamical
calculations \cite{MLP98a}.
 \end{small}}
 \end{minipage}
 \end{center}
 \end{figure*}

The sonic-point model explains naturally
why the separation between the
frequencies of the two kilohertz QPOs is
nearly constant and equal to the burst
oscillation frequency or half this
frequency. It is also consistent with
the accretion rates and stellar magnetic
fields inferred previously and accounts
for the main features of the kilohertz
QPOs, including their high and variable
frequencies, their high amplitudes and
coherences, and the common occurrence of
two simultaneous kilohertz
QPOs \cite{MLP98a}.

\section{Constraints from Kilohertz QPOs}

\subsection{Nonrotating Stars}

In order to see how constraints on the
equation of state of neutron star matter
can be derived, suppose first that the
star is not rotating and assume that,
for the star in question, $\nu_{\rm
QPO2}^\ast$---the highest observed value
of the frequency of the higher-frequency
(Keplerian-frequency) QPO in the
kilohertz QPO pair---is 1220~Hz (this is
the highest QPO frequency detected so
far from any neutron star; see
\cite{MLP98a}). Obviously, the orbital
radius $R_{\rm orb}$ of the clumps
producing the QPO must be greater than
the stellar radius; $R_{\rm orb}$ must
also be greater than the radius $R_{\rm
ms}$ of the innermost stable circular
orbit in order for the clumps to produce
a wave train that lasts tens of
oscillation periods, as observed. These
requirements constrain the
representative point of the star to lie
in a pie-slice shaped region of the
radius-mass plane (see Fig.~3a). This
bounds the mass and radius of the star
from above. In terms of $\nu_{\rm
QPO2}^\ast$, these bounds are
\cite{MLP98a}
 \begin{equation}
 \label{NonRotBounds}
  M^\nonrot_{\rm max} =
  2.2\,(1000\,{\rm Hz}/\nu_{\rm
  QPO2}^\ast)\;M_\odot
 \ \
 {\rm and}
 \ \
  R^\nonrot_{\rm max} =
  19.5\,(1000\,{\rm Hz}/\nu_{\rm
  QPO2}^\ast)\;{\rm km.}
 \end{equation}
 Figure~3b compares the regions of the
radius-mass plane allowed for three
values of $\nu_{\rm QPO2}^\ast$ with the
mass-radius relations for nonrotating
stars given by five representative
equations of state.

 \begin{figure*} 
 \begin{center}
 \begin{minipage}{13cm}
 \vglue-1truecm
 \centerline{\hglue-0.3truecm
\psfig{file=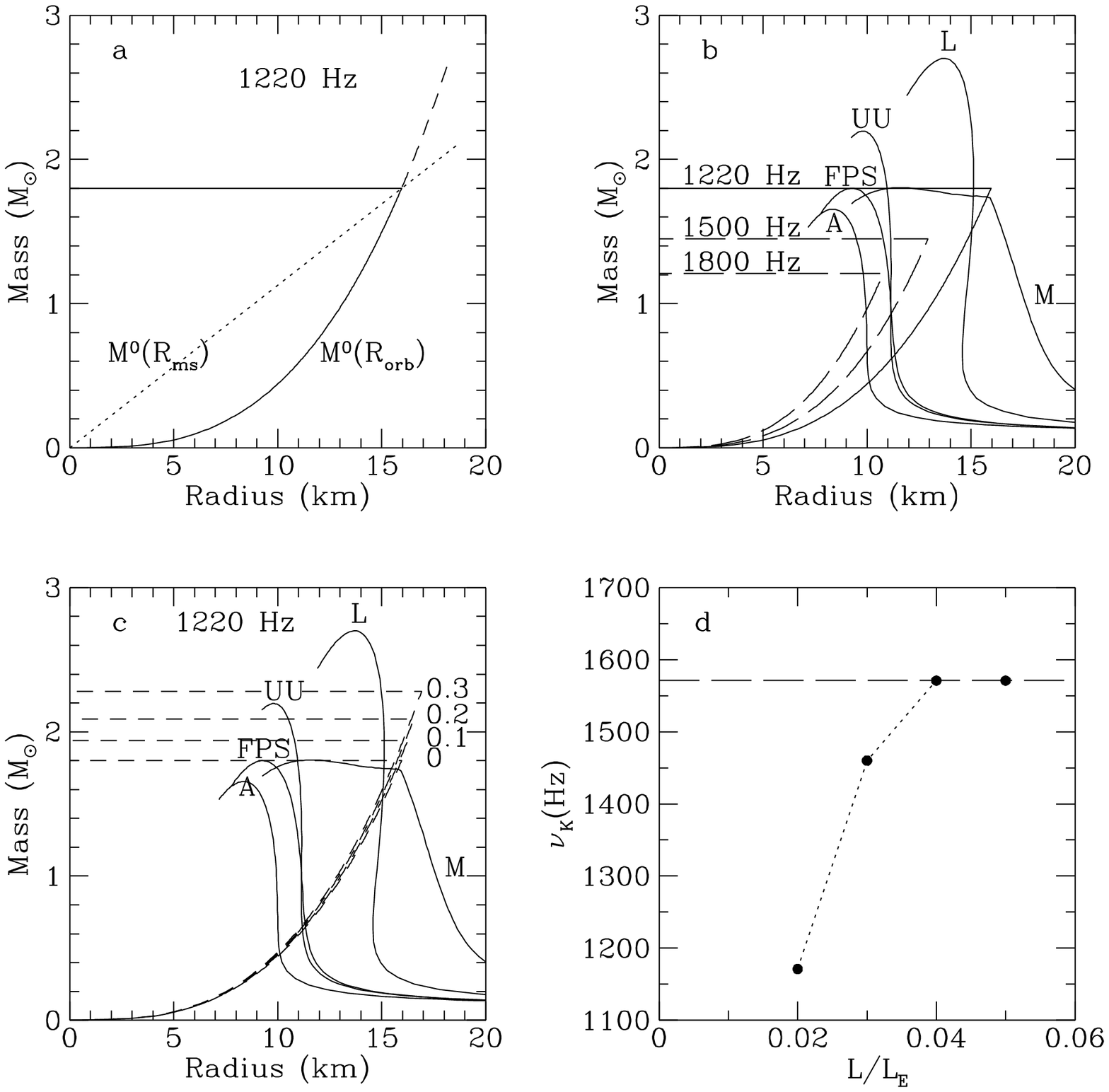,width=14truecm}
 }
 \vglue-0.25truecm
 \baselineskip=12pt
 {\begin{small}
 Fig.~3.
 (a)~Radius-mass plane, showing how to
construct the region allowed for a
nonrotating neutron star with $\nu_{\rm
QPO2}^\ast = 1220$~Hz. $R_{\rm orb}$
must be greater than the stellar radius,
so the star's representative point must
lie to the left of the (dashed) cubic
curve $M^\nonrot(R_{\rm orb})$ that
relates the star's mass to the radius of
orbits with frequency 1220~Hz.
 The high coherence of the oscillations
constrains $R_{\rm orb}$ to be
greater than $R_{\rm ms}$, the radius of
the innermost stable orbit,
which means that the radius of the 
actual orbit must lie on the
$M^\nonrot(R_{\rm orb})$ curve
below its intersection with the (dotted)
straight line $M^\nonrot(R_{\rm ms})$ that
relates the star's mass to $R_{\rm ms}$.
These requirements constrain the star's
representative point to lie in the
pie-slice shaped region enclosed by the
solid line.
  (b)~Comparison of the regions allowed
for nonrotating stars with three
different QPO frequencies with the
mass-radius relations for nonrotating
neutron stars given by five
representative equations of state.
  (c)~Regions allowed for rotating
neutron stars with four values of $j$
and $\nu_{\rm QPO2}^\ast = 1220$~Hz, when
effects of the stellar spin are included
to first-order (see text).
  (d)~Illustrative Keplerian QPO
frequency vs.\ accretion luminosity curve
predicted by the sonic-point model.
 \end{small}}
 \end{minipage}
 \end{center}
 \end{figure*}

\subsection{Rotating Stars}

Rotation affects the structure of the
star and the spacetime, altering the
region of the radius-mass plane allowed
by a given value of $\nu_{\rm
QPO2}^\ast$. Slow rotation expands the
allowed region whereas rapid rotation
shrinks it greatly. The parameter that
characterizes the importance of
rotational effects is the dimensionless
quantity \hbox{$j \equiv cJ/GM^2$},
where $J$ and $M$ are the angular
momentum and gravitational mass of the
star. For the spin frequencies
$\sim$300~\Hz inferred for the neutron
stars in the kilohertz QPO sources, $j$
is $\sim 0.1 \dash 0.3$, depending on
the mass of the star and the equation of
state. For such small values of $j$, the
structure of the star is almost
unaffected and a treatment that is
first-order in $j$ is adequate. To this
order, the existence of upper bounds on
the mass and radius can be proved
analytically \cite{MLP98a}. The bounds
for prograde orbits are
 \begin{equation}
 \label{RotBounds}
   M_{\rm max} \approx
   [1+0.75j(\nu_{\rm spin})]M^\nonrot_{\rm
max}
 \quad
 {\rm and}
 \quad
   R_{\rm max} \approx
   [1+0.20j(\nu_{\rm spin})]R^\nonrot_{\rm
max}\;,
 \end{equation}
 where $j(\nu_{\rm spin})$ is the value
of $j$ for the observed stellar spin rate
{\em at the maximum allowed mass for the
equation of state being considered\/} and
$M^\nonrot_{\rm max}$ and $R^\nonrot_{\rm
max}$ are the bounds on the mass and
radius for a nonrotating star (see
eqs.~[1]). Equations~(2) show that the
region allowed for a slowly rotating
star is always larger than the region
allowed for a nonrotating star,
regardless of the equation of state.
Figure~1c illustrates the effects of
slow stellar rotation on the allowed
region of the radius-mass plane. Detailed
calculations show that the mass of the
neutron star in \fu{1636$-$536} must be
less than $2.2\,M_\odot$ and its radius
must be less than 17~km. The upper
bounds may be smaller, depending on the
equation of state assumed
\cite{MLP98a,MLP98b}.

 \begin{figure*} [t!]  
 \begin{center}
 \begin{minipage}{13cm}
 \vglue-0.8truecm
 \centerline{\hglue+0.7truecm
\psfig{file=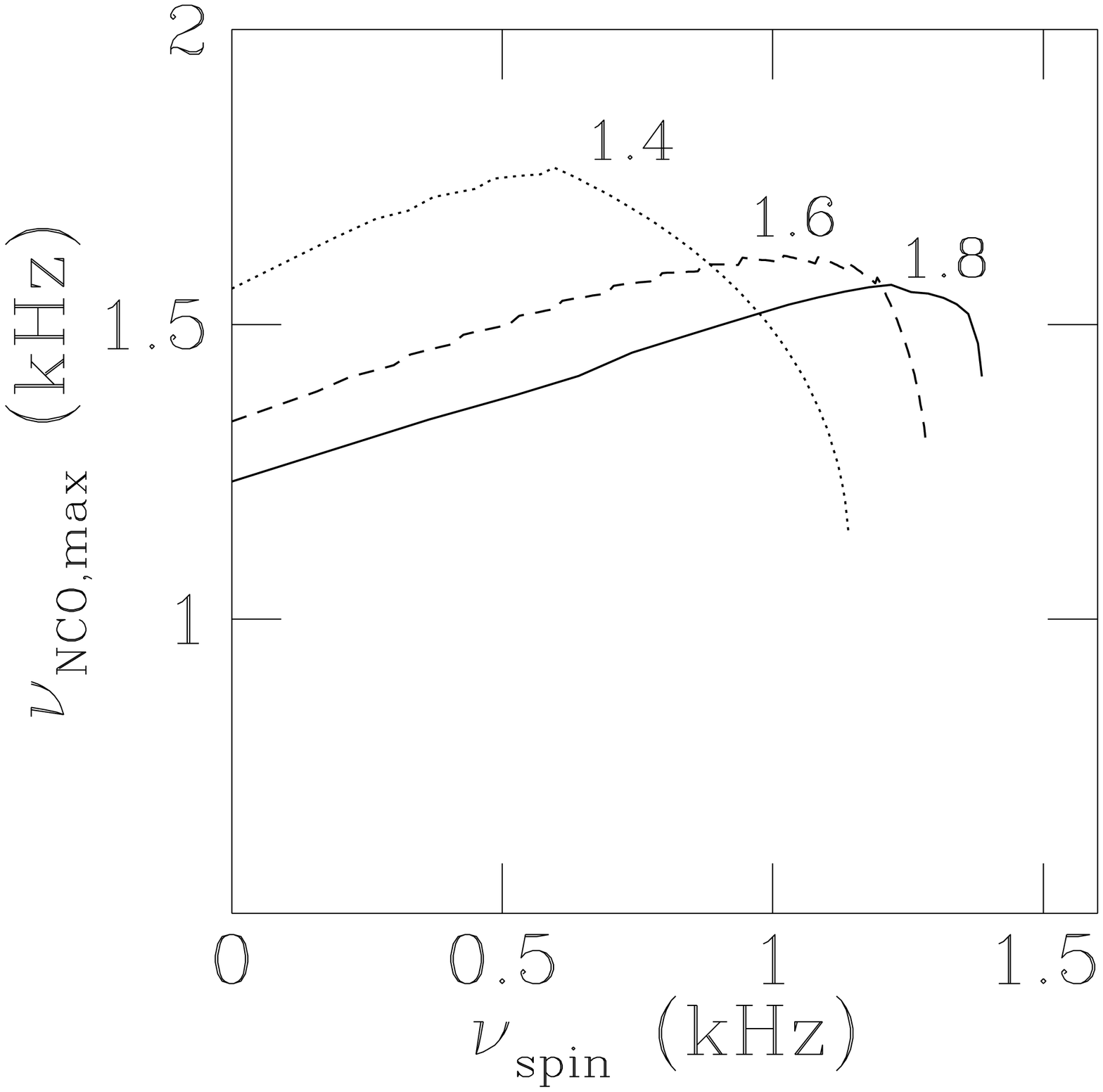,width=7truecm}
 \hglue-0.6truecm
\psfig{file=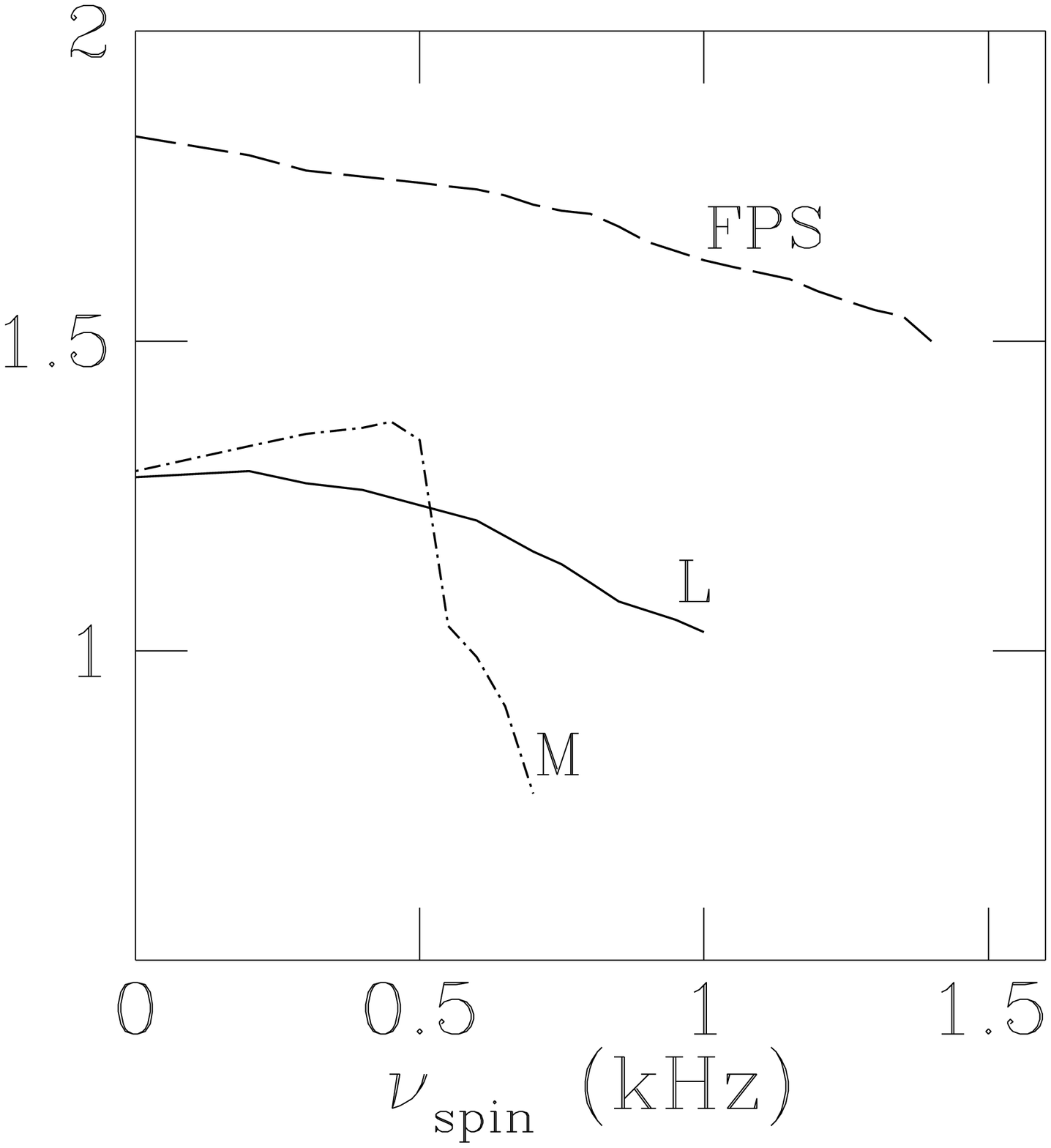,width=7truecm}
 \hglue+0.1truecm}
 \vglue+0.3truecm
 \baselineskip=12pt
 {\begin{small}
 Fig.~4.
 (a)~Maximum frequency of a nearly
circular orbit as a function of stellar
spin frequency for equation of state FPS
and stars with $1.4\,M_\odot$ (dotted
line), $1.6\,M_\odot$ (dashed line), and
$1.8\,M_\odot$ (solid line)
gravitational masses.
 (b)~Maximum frequency of a nearly
circular orbit for a star of {\it any}
mass, for the three equations of state
indicated.
 See \cite{MLC98} for details and further
discussion.
 \end{small}}
 \end{minipage}
 \end{center}
 \end{figure*}

For spin frequencies $\gta 400~\Hz$, the
structure of the star can be
significantly affected, as well as the
exterior spacetime. In this case both the
structure of the star and the spacetime
must be computed numerically for each
assumed equation of state. Determination
of the bounds on the mass and radius of a
star for a given QPO frequency and
equation of state therefore requires
construction of a sequence of stellar
models and spacetimes for different
masses using the equation of state, with
$\nu_{\rm spin}$ as measured at infinity
held fixed \cite{MLP98a}.
 Such sequences have been constructed
\cite{MLC98} and show that if the neutron
star is spinning rapidly, the constraints
on the equation of state are tightened
dramatically. For example, if the
$\sim$580~Hz burst oscillation frequency
observed in \fu{1636$-$536} is its spin
frequency (rather than twice its spin
frequency as indicated by the frequency
separation of its two kilohertz QPOs),
then very stiff equations of state like
the tensor-interaction equation of state
\cite{PS75a} are excluded by the 1220~Hz
QPO already observed from this star.
Regardless of the star's spin rate, a
1500~Hz QPO frequency would constrain
the mass and radius of the neutron star
to be less than $\sim$1.7$\,M_\odot$ and
$\sim$13~km, ruling out several
equations of state that are currently
astrophysically viable \cite{MLC98}.

Figure~4a shows why slow rotation
typically loosens the constraints on
mass and radius implied by a given QPO
frequency, whereas rapid rotation
tightens them. At low spin rates the
equatorial radius of the star is smaller
than the radius of the innermost stable
orbit and hence the maximum orbital
frequency increases linearly with the
star's spin rate (see eqs.~[2]). In
contrast, at high spin rates the
equatorial radius becomes larger than
the radius of the circular orbit that
would be marginally stable, so no
marginally stable orbit exists. The
highest frequency orbit for a star of
given mass is then the one just above the
stellar surface, which increases in
radius as the star spins faster, causing
the highest possible orbital frequency
to decrease with increasing spin rate.
Figure~4b displays the maximum frequency
of a nearly circular orbit for a star of
{\em any\/} mass, for three different
equations of state. Hence, if {\em
any\/} neutron star is found to have
spin and QPO frequencies that place its
representative point above one of these
curves, that equation of state is
excluded.

\subsection{Innermost Stable Circular
Orbit}

Establishing that an observed QPO
frequency is the orbital frequency of the
innermost stable circular orbit in an
X-ray source would be an important step
forward in our understanding of
strong-field gravity and the properties
of dense matter, because it would be the
first confirmation of a prediction of
general relativity in the strong-field
regime and would also fix (for each
equation of state) the mass of the
neutron star involved.

Given the fundamental significance of the
detection of an innermost stable orbit,
it is very important to establish what
would constitute strong, rather than
merely suggestive, evidence that the
innermost stable orbit has been detected.
 Probably the most convincing signature
would be a kilohertz QPO with a
frequency that reproducibly increases
steeply with increasing accretion rate
but then becomes constant and remains
nearly constant as the accretion rate
increases further. This behavior emerges
naturally from general relativistic
calculations of the gas dynamics and
radiation transport in the sonic-point
model (see Fig.~3d). The constant
frequency should always be the same in a
given source.
 Two other possible signatures of the
innermost stable orbit are discussed in
\cite{MLP98a}.

Several authors have recently suggested
that innermost stable orbits have already
been observed. Zhang et al.\
\cite{ZSS97} suggested that the
similarity of the highest QPO
frequencies seen so far indicates that
innermost stable orbits are being
detected and that, based on the
equations~(1) for nonrotating stars, the
neutron stars in {\em all\/} the
kilohertz QPO sources therefore have
masses close to $2.0\,M_\odot$. Kaaret,
Ford, \& Chen \cite{KFC97} suggested
that the $\sim$800--900~Hz QPOs
discovered in \fu{1608$-$52}
\cite{Berger96} and \fu{1636$-$536}
\cite{Zhang96}, which were initially
observed to have roughly constant
frequencies, are generated by the beat
of the spin frequency against the
frequency of innermost stable orbits in
these sources. However, {\em no clear
signature of an innermost stable
circular orbit has so far been seen in
any source}.

Indeed, subsequent observations of both
\fu{1608$-$52} \cite{Mendez98} and
\fu{1636$-$536} \cite{W97} are
inconsistent with the interpretation that
the frequencies of the QPOs seen
initially are related to the frequencies
of innermost stable orbits around these
stars. The 1171~Hz QPO seen by Zhang et
al.\ \cite{Zhang96}, which was assumed
by Kaaret et al.\ \cite{KFC97} to be at
the frequency of the innermost stable
orbit in \fu{1636$-$536} in order to
estimate the mass of the star, was
later seen at 1193~Hz \cite{W97} and
still later at 1220~Hz (W.\ Zhang,
personal communication). Hence, there is
as yet no evidence for a maximum QPO
frequency in \fu{1636$-$536} and hence
there is no basis for the suggestion
that an innermost stable orbit has been
seen in this source. A recent analysis of
\fu{1608$-$52} data by M\'endez et al.\
\cite{Mendez98} shows that this source
has two kilohertz QPOs that vary with
countrate just like the other sources.
There is as yet no evidence for a
maximum QPO frequency in \fu{1608$-$52}
and hence there is no basis for the
suggestion that an innermost stable
orbit has been seen in this source,
either.

\section{Constraints from Burst
Oscillations}

As noted in \S2, the strong (rms
amplitudes up to at least 35\%),
high-frequency oscillations seen during
X-ray bursts are thought to be caused by
emission from a single or two nearly
antipodal bright regions on the
stellar surface, which produce large
amplitude brightness oscillations at the
stellar spin frequency or its first
overtone as the star turns. The 
anisotropy at infinity of the radiation
emitted from such regions, and hence the
amplitude of the burst oscillation,
typically decreases with increasing
gravitational light deflection by the
star. Hence, in addition to any
constraints on the radius of the star
that may be derived from the X-ray
spectra of the bursts, constraints on the
compactness of the star (and hence the
softness of neutron star matter) can be
derived from the observed amplitudes of
the burst oscillations
\cite{ML98,Strohmayer97}. Such
constraints are particularly useful
because they complement the constraints
derived from the kilohertz QPOs, which
constrain the stiffness of neutron star
matter.

 \begin{figure*}
 \begin{center}
 \begin{minipage}{13cm}
 \vglue-0.96truecm
 \centerline{\hglue+0.45truecm
\psfig{file=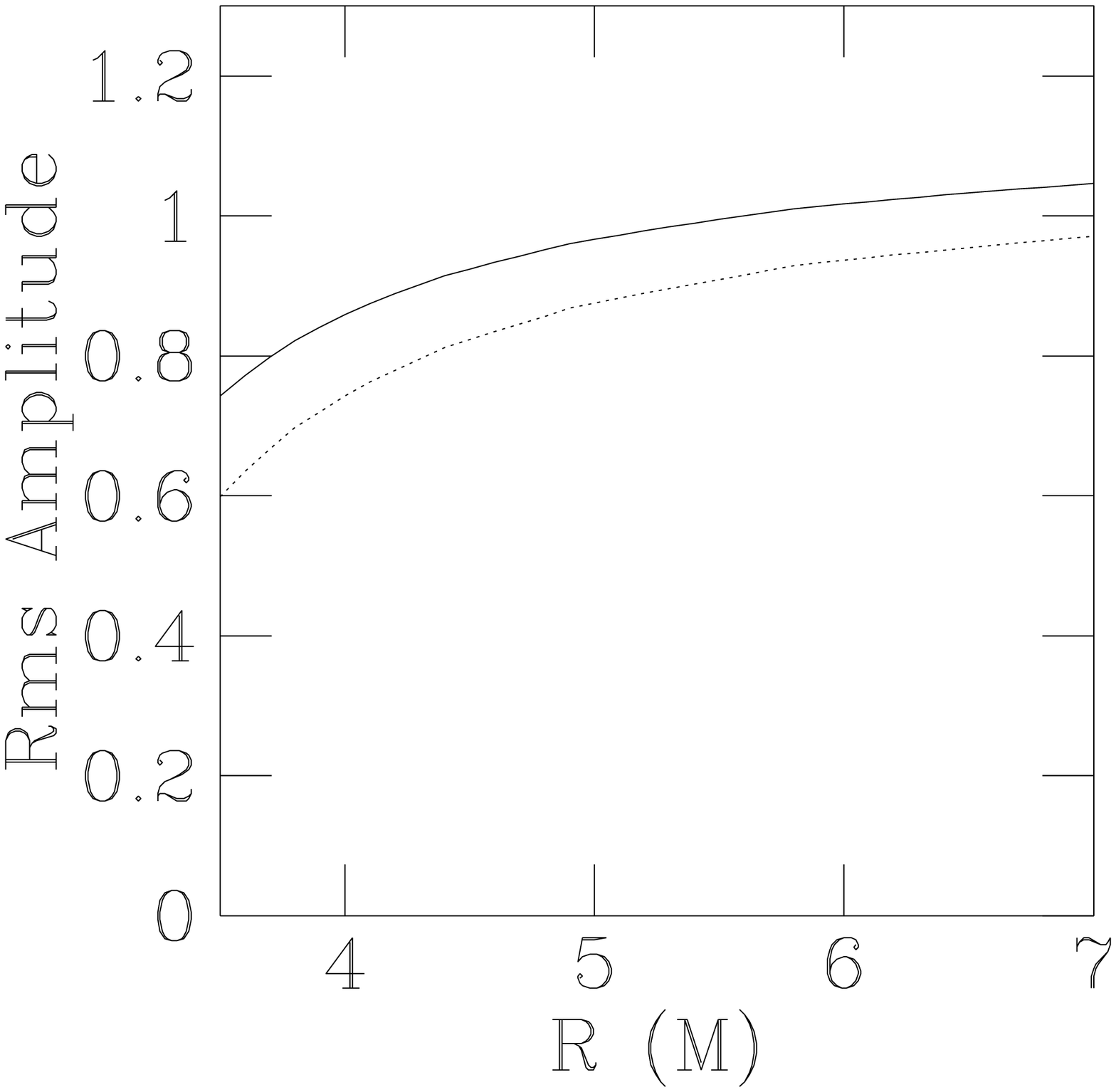,width=6.5truecm}
 \hglue-0.3truecm
\psfig{file=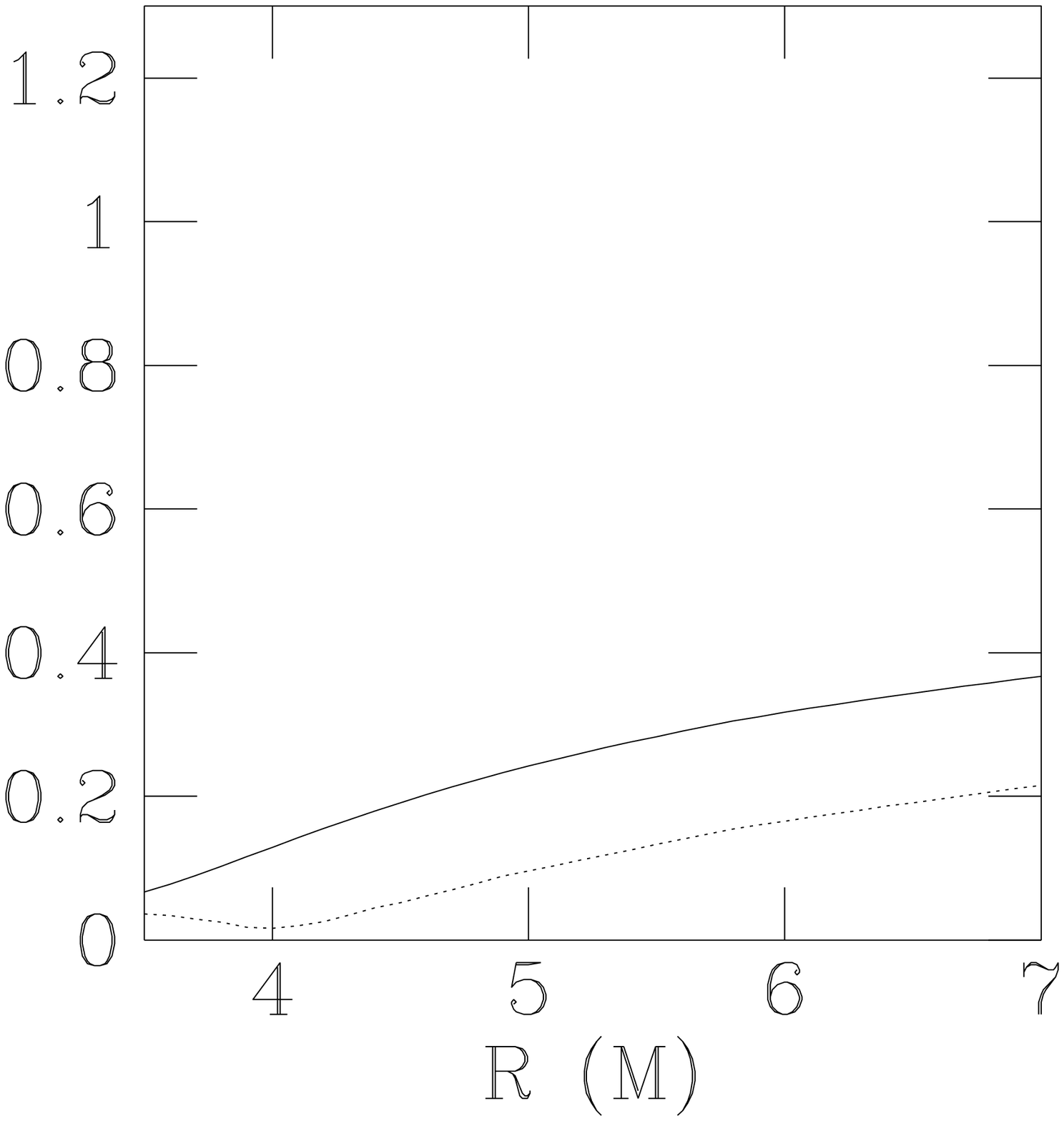,width=6.5truecm}
 }
 \vglue+0truecm
 \baselineskip=12pt
 {\begin{small}
 Fig.~5.
 Upper bounds on the observed fractional
rms amplitude of oscillations in the
photon number flux during bursts, as a
function of neutron star radius. The
observer's line of sight is assumed to
be in the rotational equator and the
aberration and Doppler shifts caused by
rotation are neglected.
 (a)~Amplitude as a function of neutron
star radius for isotropic emission
(dotted line) or the peaked emission
expected for a scattering atmosphere
(solid line) from a single point.
 (b)~Amplitude as a function of radius for
the same intensity distributions as in
(a), but for emission from two identical,
antipodal emitting points.
 See \cite{ML98} for details of the
computational method and further
discussion.
 \end{small}}
 \end{minipage}
 \end{center}
 \end{figure*}

Figure~5 shows results typical of general
relativistic calculations of the maximum
amplitudes of burst oscillations from one
or two bright regions, as a function of
stellar radius. {\em Photon counting
rate\/} oscillations with 2--60~keV rms
amplitudes as high as $\sim$35\% have
been reported in \hbox{4U~1636$-$536}
with a frequency equal to twice the
apparent spin frequency of this neutron
star. Figure~5b shows that an oscillation
this large in the {\em photon number
flux\/} would constrain the radius of
this star to be greater than $5.5\,M$.
However, the amplitude of the countrate
oscillation measured at infinity can be
larger or smaller than the amplitude of
the oscillation in the photon flux at
the star, depending on the detector
response as well as the angular
dependence and spectrum of the emission
from the stellar surface, the stellar
spin rate, and whether there is any
scattering material surrounding the
star, so more detailed modeling will be
required to extract the actual
constraints on the compactness (see
\cite{ML98}).

\section{Concluding Remarks}

The discovery using the {\em Rossi X-Ray
Timing Explorer\/} that many neutron
stars with weak magnetic fields produce
strong $\sim$300--1200~Hz X-ray
brightness oscillations is a spectacular
achievement that validates both the
scientific expectations that led to the
mission and the long years of hard work
that were needed to bring it to fruition.
The kilohertz QPOs discovered in the
accretion-powered emission are already
providing interesting new upper bounds on
the masses and radii of neutron stars,
and on the stiffness of neutron star
matter. The high-frequency oscillations
discovered in the emission during
thermonuclear X-ray bursts are likely to
provide interesting new bounds on the
compactness of neutron stars and hence
on the softness of neutron star matter.
Observation of a QPO with a frequency
just 100~Hz higher than the highest
frequency so far seen would exclude the
stiffest proposed neutron star matter
equations of state. 

Observation of innermost stable circular
orbits would be the first confirmation
of a strong-field prediction of general
relativity and  would fix the mass of
the star involved, for each equation of
state considered. Although there is
currently no strong evidence that an
innermost stable circular orbit has been
discovered around any of these neutron
stars, there is reason to hope that such
evidence may be forthcoming. Given the
rapid pace of discoveries with \rxte,
the prospects for obtaining compelling
evidence of an innermost stable circular
orbit appear good.

\medskip

This work was supported in part by NSF
grant AST~96-18524, NASA grant
NAG~5-2925, and NASA RXTE grants at the
University of Illinois, and by NASA
grant NAG~5-2868 at the University of
Chicago.


\begin{thebibliography}{99} \itemsep=0cm


\bibitem{Berger96}
 Berger, M., et al.\ 1996, ApJ, 469, L13

\bibitem{ELP86}
 Epstein, R., Lamb, F.\,K., \&
Priedhorsky, W. 1986, Astrophysics of
Time Variability in X-Ray and Gamma-Ray
Sources, Los Alamos Science, No.~13

\bibitem{KFC97}
 Kaaret, P., Ford, E.~C., \& Chen, K.
1997, ApJ, 480, L27

\bibitem{Klein96}
 Klein, R.\,I., Jernigan, J.\,G., Arons,
J., Morgan, E.\,H., \& Zhang, W. 1996,
ApJ, 469, L119


\bibitem{Lamb91}
 Lamb, F.\,K. 1991, in Neutron Stars:
Theory and Observation, ed. J. Ventura \&
D. Pines (Dordrecht: Kluwer), 445


\bibitem{LP79}
 Lamb, F.\,K., \& Pines, D. 1979,
Compact Galactic X-Ray Sources (Urbana:
Univ. of Illinois Physics Dept.)



\bibitem{McClintock98}
 McClintock, J.\,E. 1998, in Accretion
Processes in Astrophysical Systems, ed.
S. Holt \& T. Kallman (AIP Conf. Proc.),
in press (astro-ph/9802080)

\bibitem{Mendez98}
 M\'endez, M., et al. 1998, ApJ, 494, L65

\bibitem{ML93}
 Miller, M.\,C., \& Lamb, F.\,K. 1993,
ApJ, 413, L43

\bibitem{ML96}
 ---------. 1996, ApJ, 470, 1033

\bibitem{ML98}
 ---------. 1998, ApJ, in press
(astro-ph/9711325)

\bibitem{MLC98}
 Miller, M.~C., Lamb, F.~K., \& Cook, G.
1998, ApJ, submitted

\bibitem{MLP98a}
 Miller, M.~C., Lamb, F.~K., \& Psaltis,
D. 1998, ApJ, in press (astro-ph/9609157)

\bibitem{MLP98b}
  ---------. 1998, ApJ, in preparation



\bibitem{PS75a}
 Pandharipande, V.~R., \& Smith, R.~A.
1975, Nucl. Phys., A237, 507


\bibitem{SMB97} Smith, D.\,A., Morgan, E.,\,H.,
\& Bradt, H. 1997, ApJ, 479, L137

\bibitem{Strohmayer97}
 Strohmayer, T.\,E., talk presented at
the 1997 November meeting of the High
Energy Astrophysics Division of the
American Astronomical Society

\bibitem{SSZ98}
 Strohmayer, T.\,E., Swank, J.\,H., \&
Zhang, W. 1998, in The Active X-Ray Sky,
eds. L.\,Scarsi, H.\,Bradt, P.\,Giommi,
and F.\,Fiore, Nucl. Phys. B Proc.
Suppl., in press (astro-ph/9801219)
 
\bibitem{SZS97} Strohmayer, T.\,E., Zhang, W., \& Swank, J.\,H.
1997, ApJ, 487, L77

\bibitem{Strohmayer96}
 Strohmayer, T., Zhang, W., Swank,
J.\,H., Smale, A., Titarchuk, L., \&
Day, C. 1996, ApJ, 469, L9

\bibitem{Swank95}
 Swank, J., et al. 1995, in The Lives of
Neutron Stars, ed. M.\,A. Alpar, \"U.
K{\i}z{\i}lo{\v g}lu, \& J. van Paradijs
(Dordrecht: Kluwer), 525

\bibitem{TM97}
 Titarchuk, L., \& Muslimov, A. 1997,
A\&A, 323, L5

\bibitem{vdK98}
 van der Klis, M. 1998, in The Many
Faces of Neutron Stars, Proc. NATO ASI,
Lipari, Italy (Dordrecht: Kluwer), in
press (astro-ph/9710016)

\bibitem{vdK97}
 van der Klis, M., Wijnands, R., Horne,
K., \& Chen, W. 1997, ApJ, 481, L97

\bibitem{W97}
 Wijnands, R.A.D., et al. 1997, ApJ,
479, L141


\bibitem{Zhang96}
 Zhang, W., Lapidus, I., White, N.~E.,
\& Titarchuk, L. 1996, ApJ, 469, L17

\bibitem{ZSS97}
 Zhang, W., Strohmayer, T., \& Swank,
J.~H. 1997, ApJ, 482, L167

\end{thebibliography}
\end{document}